\documentclass[%
 reprint,
superscriptaddress,
 amsmath,amssymb,
 aps,
prl,
longbibliography
]{revtex4-2}

\usepackage{graphicx}
\usepackage{dcolumn}
\usepackage{bm}

\begin{document}

\title{ Stress tensor field and mesoscopic stresses in the vertex model for tissues }

\author{Paulo C. Godolphim}
\affiliation{Departamento de F\'\i sica, FCFM, Universidad de Chile, Santiago, Chile}
\affiliation{Instituto de Física, Universidade Federal do Rio Grande do Sul, Porto Alegre, Brasil.}

\author{Leonardo G. Brunnet}
\affiliation{Instituto de Física, Universidade Federal do Rio Grande do Sul, Porto Alegre, Brasil.}

\author{Rodrigo Soto}%
\affiliation{Departamento de F\'\i sica, FCFM, Universidad de Chile, Santiago, Chile}

\date{\today}

\begin{abstract} 
Mechanical stresses are fundamental regulators in biological tissues, where the vertex model (VM) is pivotal for theoretical and force-inference studies. Yet, no uniform expression for the stress tensor exists for the VM. Here we provide a microscopic derivation of it, linking mesoscopic stresses to the VM forces.  The stress field presents a freedom on how tensions are distributed across cells, which allows previous expressions to emerge as particular realizations of the field and suggests a link between mesoscopic stresses and cytoskeletal force-transmission architectures in real cells.
\end{abstract}

\maketitle

\textit{Introduction.} Mechanical forces and stresses play a central role in tissue development, homeostasis, and disease progression~\cite{Heisenberg2013,Goodwin2021,Di2023}. Mechanical stresses orient cell divisions during zebrafish epiboly~\cite{Campinho2013}, regulate stem-cell differentiation under geometrical confinement and stretching~\cite{McBeath2004}, drive tissue invagination during \textit{Drosophila} gastrulation~\cite{Keller2003}, and generate large-scale tissue flows during primitive streak formation in chicken embryos~\cite{Saadaoui2020,Caldarelli2024}.
Understanding how forces are generated, transmitted, and organized into tissue-scale stresses is therefore a central problem in tissue mechanics.

Among the theoretical frameworks used to study tissue mechanics, the vertex model (VM) is one of the most widely employed~\cite{Fletcher2014,Alt2017,Lange2025}. It has been used to describe epithelial dynamics across many biological systems, including zebrafish retinal organization~\cite{Salbreux2012}, \textit{Fundulus} epiboly~\cite{Weliky1990}, killifish epiboly~\cite{Verdugo2022}, \textit{Xenopus} notochord development~\cite{Weliky1991}, mouse blastocyst polarity~\cite{Honda2008}, \textit{Drosophila} wing development~\cite{Mao2013,Sui2018,Tetley2019}, while capturing key qualitative behaviors such as rigidity transitions and cell rearrangements~\cite{Farhadifar2007,Staple2010,Bi2014,Bi2015,Arzash2025}. 
Moreover, it serves as the basis for a broader class of active-vertex-~\cite{Barton2017}, viscoelastic-~\cite{Verdugo2022}, Voronoi-~\cite{Bi2016,Yang2017}, and network-based~\cite{Noll2017,Kim2021,Dow2023} models used to study collective cell mechanics, and, more recently, to develop mechanosensitive models that couple force transmission to cytoskeletal regulation and mechanotransduction~\cite{Noll2017,Naik2026,Lange2025}. 

The VM also serves as a mathematical basis of many force-inference approaches used to extract mechanical information from experimental systems~\cite{Ishihara2012,Chiou2012,Brodland2014,Roffay2021}. Although stresses and mechanical properties can be experimentally accessed through techniques such as traction-force microscopy~\cite{Plotnikov2014,Ribeiro2016}, atomic-force microscopy~\cite{Schierbaum2019}, micropipette aspiration~\cite{Guevorkian2017}, laser ablation~\cite{Kong2019}, embedded magnetic and oil droplets~\cite{Campas2014,Serwane2016,Lucio2017}, and tissue stretching assays~\cite{Trepat2004,Garcia2020,Roshanzadeh2020,Hart2021}, these approaches can be expensive, invasive, or experimentally challenging, particularly in \textit{in vivo} systems~\cite{Dow2023}. Force-inference methods have emerged as an alternative, allowing mechanical information to be extracted directly from experimental images~\cite{Ishihara2012,Chiou2012,Brodland2014,Kong2019,Vanslambrouckid2024,Jurado2025,Roffay2021}, enabling inference of stress distributions during tissue morphogenesis~\cite{Ishihara2012} and revealing correlations between mechanical anisotropies and cell differentiation~\cite{Chiou2012}. As many of these approaches rely on the VM, their mechanical interpretation depends on how stresses are represented within the model.

Despite its relevance and widespread use in tissue mechanics,  the
precise way to obtain stresses in the VM remains unsettled and different formulations for the stress tensor have been proposed~\cite{Ishihara2012,Yang2017,Barton2017,Nestor-Bergmann2018,Jensen2020,Perez-VerdugoThesis2021,Tong2022,Jensen2023}. These expressions often yield different mesoscopic results and lack direct microscopic derivations from the VM dynamics, in the sense that they are not explicit derivations of the VM Cauchy stress~\cite{Admal2010}. Consequently, the relation between existing stress tensors and the underlying VM mechanics remains unclear. This ambiguity is not merely a mathematical issue; it directly leads to different mechanical interpretations of the same tissue, while the absence of a microscopic derivation makes it difficult to determine which interpretation is more appropriate. Furthermore, the VM exhibits a parametric degeneracy~\cite{Yang2017,Godolphim2025}, which can be identified as a pressure-gauge freedom~\cite{Godolphim2025}. As a consequence, stresses are defined only up to an additive pressure constant, an aspect that must be considered in the formulation of the stress tensor.

In this Letter, we derive a stress field for the VM, $\boldsymbol{\sigma}(\mathbf{r})$, which is invariant under the pressure-gauge degeneracy and, when integrated over regions surrounding the vertices, correctly reproduces the forces acting on them, thereby recovering the microscopic VM dynamics. This construction establishes the missing bridge between microscopic force transmission and mesoscopic stress measurements, yielding explicit expressions for tissue-, cell-, edge-, and vertex-level stress tensors and enabling a direct comparison with existing formulations. In molecular systems, the stress tensor associated with a given set of particle forces is not uniquely defined~\cite{IrvingKirkwood1950,soto2016kinetic}. We show that an analogous situation arises in the VM, where previously proposed stress tensors emerge as different realizations of the same underlying force field rather than as competing descriptions. Furthermore, this freedom admits a direct biophysical interpretation, associated with the manner in which the edge forces are effectively transmitted by cytoskeleton \textit{bundle fibers}. Lastly, the framework presented here holds for irregular, regular, heterogeneous, homogeneous, open, and closed tissues.

\textit{The vertex model.}
In the VM, cellular tissue is modeled as a two-dimensional tiling of polygonal cells that share edges and vertices. The state of the tissue is  given completely by the position $\mathbf{r}_i$ of the vertices, which move following a variational dynamics
$\dot{\mathbf{r}}_i=\mathbf{f}_i=-\frac{\partial E}{\partial \mathbf{r}_i}$, where $E$ is the tissue energy functional, which depends on the vertex positions through the cell areas and edge lengths. The results presented in this Letter do not assume any particular form of $E$, but whenever concrete expressions are needed, we consider the area-perimeter form
$
    E = \sum_c \left [ \frac{K_{Ac}}{2}\left (A_c - A_{0c} \right )^2 + \frac{K_{Lc}}{2}(L_c - L_{0c})^2 \right],
$
which penalizes deviations of the cell areas and perimeters from their target values. The sum is over all cells $c$, $A_c$ and $A_{0c}$ denote the actual and target cell areas, $L_c$ and $L_{0c}$ denote the actual and target cell perimeters, and $K_{Ac}$ and $K_{Lc}$ are the area and perimeter stiffness controlling the relaxation of cells toward their target values.  
Here, for generality, we consider a heterogeneous formulation in which parameters vary from cell to cell. The homogeneous VM, which is a common approximation, consists of taking $A_{0c}=A_0$, $L_{0c}=L_0$, $K_{Ac}=K_A$, and $K_{Lc}=K_L$.
The VM allows to define the cell pressure $P_c \equiv -\frac{\partial E}{\partial A_c} = - K_{Ac} (A_c - A_{0c})$ and the edge tension $T_{e} \equiv \frac{\partial E}{\partial l_e} = \sum_{c / e\in c}K_{Lc} \left ( L_{c} - L_{0c} \right )$, where $l_e=|\mathbf{l}_e|$ is the edge length, and the sum is over the two neighbor cells whose shared membrane forms the edge $e$~\cite{Chiou2012}. The parameter degeneracy in the VM consists in that modifying the value of $\sum_c K_{Ac}A_{0c}$, which changes the average pressure, does alter the system dynamics~\cite{Yang2017,Godolphim2025}.

\begin{figure}[htb]
    \centering
    \includegraphics[width=\linewidth,trim=25 477 25 40, clip]{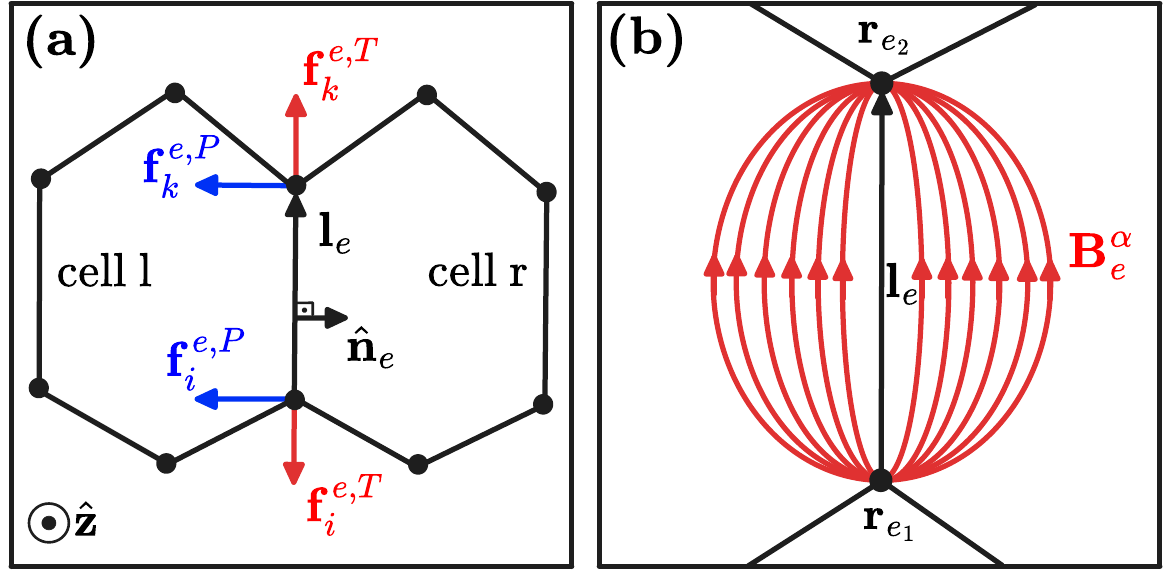}
    \caption{
    (a) Force decomposition at edge $e$. Cells lie in the plane with normal $\hat{\mathbf z}$. The pressure contribution (blue) acts normally to the edge and depends on the pressure difference between neighboring cells, while the tension contribution (red) acts tangentially along the edge orientation $\hat{\mathbf l}_e$. The resulting forces on the vertices $i$ and $k$ are given by Eq.~(1). 
    (b) Schematic representation of the spread of the edge tension inside the tissue through a bundle of fibers $\mathbf B_e^\alpha$ connecting the edge vertices $\mathbf r_{e_1}$ and $\mathbf r_{e_2}$. Different fibers represent possible paths through which the edge force can be transmitted within the cells, motivating the construction of a continuous tension stress field.
    }
    \label{fig-1}
\end{figure}

The force $\mathbf{f}_i$ on a vertex can be indistinctly expressed as the sum of the forces from the cells containing the vertex, $\mathbf{f}_i^c$, or as the sum of the forces from the edges meeting at the vertex, $\mathbf{f}_i^e$ (see Supplemental Material). To derive the stress tensor, we consider the edge formulation. Separating the pressure and tension contributions, $\mathbf{f}_i^e=\mathbf{f}_i^{e,P} + \mathbf{f}_i^{e,T}$, one obtains ~\cite{Chiou2012}
\begin{align}
\mathbf{f}_i^{e,P} &= \frac{1}{2} \left ( P_{e}^{\text{l}} - P_{e}^{\text{r}} \right )l_e \hat{\mathbf{n}}_{e}, &
 \mathbf{f}_i^{e,T} &= T_{e}\hat{\mathbf{l}}_{e},\label{force_edge_pressure_tension}
\end{align}
where $P_{e}^{\text{l,r}}$ are the cell pressures at the left or right side of the edge oriented along $\mathbf{l}_e$, and  $\hat{\mathbf{n}}_{e}\equiv \hat{\mathbf{l}}_{e}\times\mathbf{\hat{z}}$ is the normal unit vector to the edge surface, with $\hat{\mathbf{z}}$ the normal vector to the plane  (Fig.~\ref{fig-1}-a).
In this formulation, the force field is expressed as a sum of pairwise forces, comprising two independent contributions: the tension component is central and reciprocal, while the pressure component is non-reciprocal.

As indicated in the Introduction, several expressions for the stress tensor have been proposed for the VM. 
The first one traces back to Ref.~\cite{Ishihara2012}. Their definition, when written for the entire tissue, reads 
$\boldsymbol{\sigma}_\mathcal{T}^\text{Ishi.} 
    = \left [ 
    - \sum_c P_cA_c \mathbf{ \mathbb{I} } + \sum_{e} T_e l_e \hat{\mathbf{l}}_e \otimes \hat{\mathbf{l}}_e
    \right ]/A_\mathcal{T}$,
where the sums are over all cells and edges, and $A_\mathcal{T}=\sum_c A_c$ is the total tissue area.
At a smaller scale, the stress on the cellular level has been proposed to be $
    \boldsymbol{\sigma}_c^\text{Yang} = - P_c \mathbf{ \mathbb{I} } + \frac{1}{2 A_c} \sum_{e \in c} T_e l_e \hat{\mathbf{l}}_e \otimes \hat{\mathbf{l}}_e
$~\cite{Yang2017}, where the  sum runs over all edges $e$ of cell $c$.
Both expressions are related by  $\boldsymbol{\sigma}_\mathcal{T}^\text{Ishi.} = \sum_c A_c\boldsymbol{\sigma}_c^\text{Yang}/A_\mathcal{T}$. Virial-like stresses at the cell scale ($\boldsymbol{\sigma}_c^\text{virial} \propto \sum_{i\in c}\mathbf{f}_i^c \otimes \mathbf{r}_i$) have also been proposed~\cite{Nestor-Bergmann2018,Jensen2020,Perez-VerdugoThesis2021,Jensen2023}. 
The specific forms of these expressions vary, but they also satisfy $\boldsymbol{\sigma}_\mathcal{T}^\text{Ishi.} = \sum_c A_c\boldsymbol{\sigma}_c^\text{virial}/A_\mathcal{T}$ (see Supplemental Material). 
Within those, Refs.~\cite{Jensen2020,Jensen2023} reported non-symmetric stresses at the vertex-level in irregular tissues, while introducing a discrete Airy stress potential. 
Other approaches include isotropic Murdoch-Hardy's constructions~\cite{Barton2017}, and provide continuous coarse-grained descriptions~\cite{Triguero-Platero2023} reproducing the VM solid--liquid transition~\cite{Bi2015}. 

\textit{Stress field.}
The objective is to construct a continuous stress field for the tissue 
$\boldsymbol{\sigma}\left( \mathbf{r}\right)=\boldsymbol{\sigma}\left( \mathbf{r};\{P_c\},\{T_{e}\} \right) = \boldsymbol{\sigma}^P\left( \mathbf{r};\{P_c\} \right) + \boldsymbol{\sigma}^T\left( \mathbf{r};\{T_{e}\} \right)$ with the following properties: 
(i)~It is degeneration-free, meaning that $\boldsymbol{\sigma}\left( \mathbf{r};\{P_c+P_0\},\{T_{e}\} \right) = \boldsymbol{\sigma}\left( \mathbf{r};\{P_c\},\{T_{e}\} \right), \forall P_0$. 
(ii)~When the induced density force $\mathbf{q}\left( \mathbf{r}\right)=\nabla \cdot \boldsymbol{\sigma}\left( \mathbf{r}\right)$ is integrated in a region $A_i$ containing only the vertex $i$, gives the total force  acting on the vertex~\cite{landau1986elasticity}:
\begin{equation}\label{eq-stress_force_vertex_stress_definition}
    \mathbf{f}_i=\int_{A_i} \nabla \cdot \boldsymbol{\sigma}\left( \mathbf{r}\right) \,d\mathbf{r}^2=\oint_{S_i}\boldsymbol{\sigma}\left( \mathbf{r}\right)\cdot\hat{\mathbf{n}}\,dS,
\end{equation}
with $S_i$ is the linear boundary of $A_i$ and $\hat{\mathbf{n}}$ is the outer normal unit vector perpendicular to the contour.
And (iii) that it can  give coarse-grained averages
\begin{equation}\label{eq-stress_mesoscopi_stress_tensor_definition}
    \boldsymbol{\sigma}_{\mathcal{X}} \equiv \left < \boldsymbol{\sigma}(\mathbf{r}) \right >_{A_\mathcal{X}} = \frac{1}{A_\mathcal{X}}\int_{A_\mathcal{X}} \boldsymbol{\sigma}(\mathbf{r})d^2\mathbf{r},
\end{equation}
where $A_\mathcal{X}$ is any area region in the tissue, e.g., a cell. These coarse-grained stresses can be used to build upscale averages, e.g., for the total tissue   $\boldsymbol{\sigma}_\mathcal{\mathcal{T}} = \sum_\mathcal{X} \boldsymbol{\sigma}_\mathcal{X} A_\mathcal{X} / A_\mathcal{T}$.
No symmetries are enforced on the stresses and they will emerge depending on the case.

As for molecular fluids, using the Irving--Kirkwood procedure~\cite{IrvingKirkwood1950,soto2016kinetic}, the force density $\mathbf{q}(\mathbf{r})\equiv \sum_i \mathbf{f}_i \delta\left ( \mathbf{r}-\mathbf{r}_i \right )$ can be written as $\mathbf{q}(\mathbf{r})= \mathbf{q}^P(\mathbf{r}) + \nabla\cdot \boldsymbol{\sigma}^{\text{micro, }T}(\mathbf{r})$, where
\begin{align} \label{eq.forcedensity.qp}
    \mathbf{q}^P(\mathbf{r})\equiv\sum_{e} \frac{P_{e}^{l} - P_{e}^{r}}{2} l_{e}\hat{\mathbf{n}}_{e}\left [ \delta(\mathbf{r}-\mathbf{r}_{e_1}) + \delta(\mathbf{r}-\mathbf{r}_{e_2}) \right ],
\end{align}
which cannot be further simplified due to the nonreciprocity of the pressure forces, and
\begin{equation}\label{eq-stress_tensio_field_micro}
    \boldsymbol{\sigma}^{\text{micro, }T}(\mathbf{r})
    \equiv \sum_{e}T_{e}\hat{\mathbf{l}}_{e}\otimes\int_{\mathbf{r}_{e_1}}^{\mathbf{r}_{e_2}}\delta(\mathbf{r}-\mathbf{b})d\mathbf{b},
\end{equation}
is the \textit{microscopic} tension field. Here, the line integral can take  \textit{any} path connecting the vertices of $e$, making, as usual, $\boldsymbol{\sigma}^{\text{micro, }T}(\mathbf{r})$ non uniquely defined~\cite{Admal2010}. Despite of this ambiguity, the total tissue tension stress is unique $\boldsymbol{\sigma}_{\mathcal{T}}^T \equiv \left < \boldsymbol{\sigma}^{\text{micro, }T}(\mathbf{r}) \right >_\mathcal{T} = \sum_{e}T_{e} l_e \hat{\mathbf{l}}_{e} \otimes \hat{\mathbf{l}}_{e} /A_\mathcal{T}$, which coincides with the tension part of $\boldsymbol{\sigma}_\mathcal{T}^\text{Ishi.}$.

By construction, $\boldsymbol{\sigma}^{\text{micro, }T}(\mathbf{r})$ fulfills Eq.~\eqref{eq-stress_force_vertex_stress_definition}. However, it is nonzero only along the path of integration. Besides being arbitrary, this makes the stress field highly singular. However, the biophysics behind the VM helps us smooth the field and resolve the ambiguity of the path. Indeed, in epithelial tissues, edge forces, which can be interpreted as cell-cell adhesion forces, are transmitted between cells through adhesions (cadherins) connected to bundles of actin filaments, which can be located both near the membrane and spread throughout the entire cell~\cite{Guillot2013,Kale2018}.
In terms of the expression \eqref{eq-stress_tensio_field_micro}, this means that there is no one but rather  
a continuous distribution of possible paths as sketched in Fig.~\ref{fig-1}-b. 
With this, the tension field can be written as
\begin{equation}\label{eq-tension_field}
    \boldsymbol{\sigma}^T(\mathbf{r}) \equiv \sum_e  T_e \, \hat{\mathbf{l}}_e  \otimes  \boldsymbol{\beta}_e(\mathbf{r}),
\end{equation}
where the \textit{spread field} $\boldsymbol{\beta}_e(\mathbf{r})\equiv \int d\alpha\, \mu_e(\alpha) \int_{\mathbf{B}_e^\alpha}\delta(\mathbf{r}-\mathbf{b})d\mathbf{b}$ encodes how tension is distributed in the tissue, where $\alpha$ label each fiber with path $\mathbf{B}_e^\alpha$, all having end points at the edge vertices $\mathbf{r}_{e_{1,2}}$, and $\mu_e(\alpha)$ are weight functions, normalized to $\int d\alpha\,\mu_e(\alpha)=1$. The spread field satisfies $\int_{A_\mathcal{T}} \boldsymbol{\beta}_e(\mathbf{r}) d^2\mathbf{r} = l_e \hat{\mathbf{l}}_e$, thus correctly reproducing the total tissue tension.

\begin{figure}[htb]
    \centering
    \includegraphics[width=\linewidth,trim=25 195 25 35, clip]{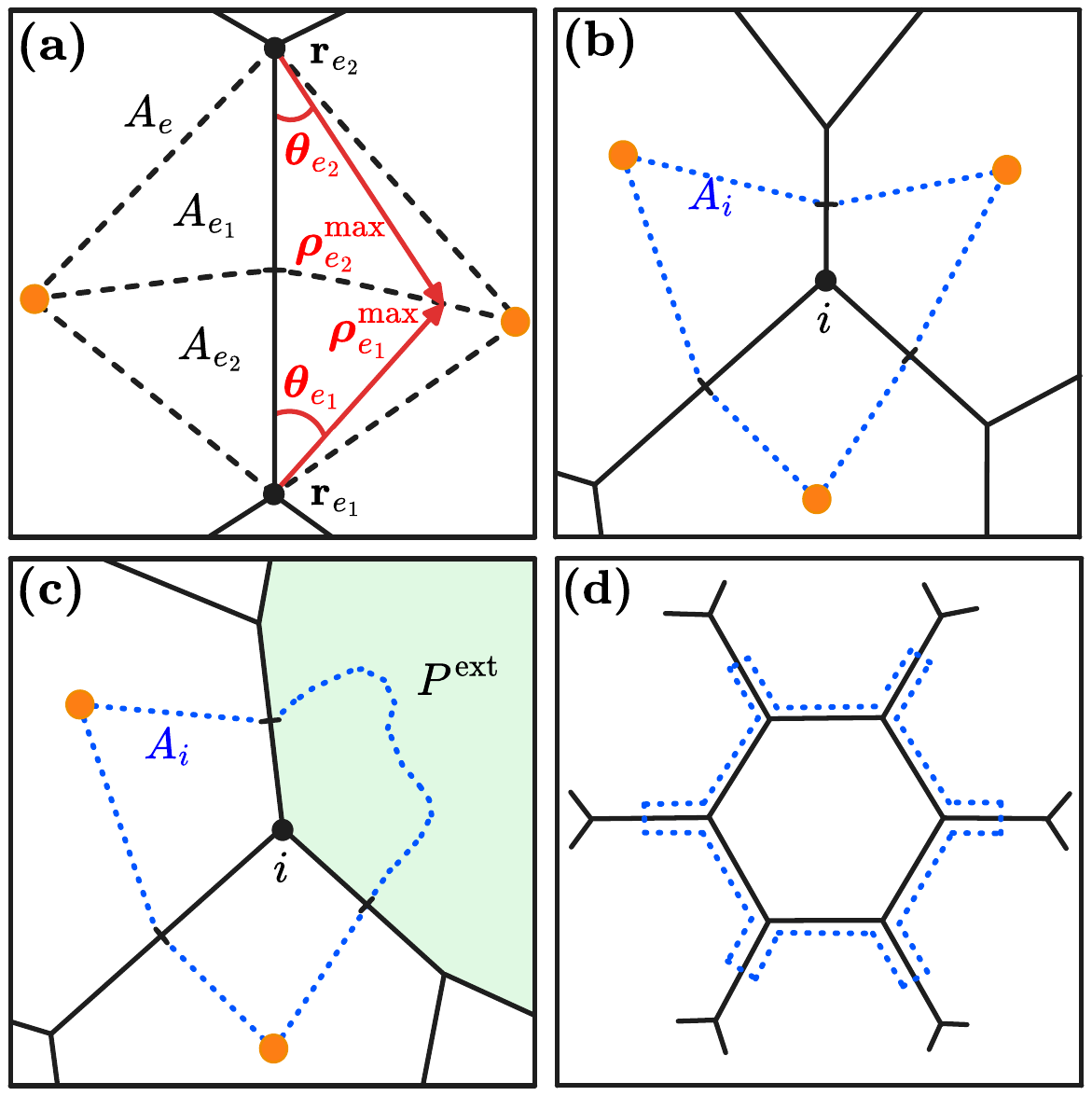}
    \caption{
    (a) Definition of the edge region $A_e$, which is divided into the subregions $A_{e_1}$ and $A_{e_2}$ associated with the vertices $\mathbf r_{e_1}$ and $\mathbf r_{e_2}$. Local polar coordinate systems $(\boldsymbol{\rho}_{e_{1,2}},\boldsymbol{\theta}_{e_{1,2}})$ are defined in each subregion, where $\boldsymbol{\rho}^{\max}_{e_{1,2}}=\boldsymbol{\rho}^{\max}_{e_{1,2}}(\theta_{e_{1,2}})$ denotes the maximum radial extent of the fibers in each region, and notice that $\boldsymbol{\rho}^{\max}_{e_{1}}(\theta_{e_{1}})-\boldsymbol{\rho}^{\max}_{e_{2}}(\theta_{e_{2}})=\mathbf{l}_e=\mathbf{r}_{e_2}-\mathbf{r}_{e_1}$, with $\boldsymbol{\rho}^{\max}_{e_{1}}(0)=-\boldsymbol{\rho}^{\max}_{e_{2}}(0)=\mathbf{l}_e/2$. Orange dots indicate the centers of the neighboring cells. (b) Vertex integration region $A_i$ associated with the vertex $i$. The dotted contour, going through the centers of the adjacent edges, defines an integration path enclosing the total force density acting on the vertex. (c) Integration contour for a border vertex in an open tissue under external pressure $P^{\mathrm{ext}}$. Outside the tissue, the contour is arbitrary, provided it passes through the centers of the border edges. (d) Example of integration contours used to compute the total force acting on a cell. The contours are taken arbitrarily close to the cell boundary while remaining outside the cell and passing through the adjacent edges' centers.
    }
    \label{fig-2}
\end{figure}

Up to now, the fiber paths and weights have been arbitrary. To obtain an explicit form of $\boldsymbol{\beta}_e(\mathbf{r})$, we consider that the fiber bundle is entirely enclosed in the region $A_e$ that is limited by the edge vertices and the centers of the two neighbor cells (Fig.~\ref{fig-2}-a). This region is further split into two: $A_{e_1}$ and $A_{e_2}$, which are limited by one of the two vertices, the cell centers, and the central point of the edge, as shown in  Fig.~\ref{fig-2}-a. With this construction, we take fibers that go radially with an orientation $\theta_{e_1}(\alpha)$ from vertex $\mathbf{r}_{e_1}$ to the line dividing the two regions and then going again radially with an orientation $\theta_{e_2}(\alpha)$ to the vertex $\mathbf{r}_{e_2}$ as shown in the Figure. This leads to
\begin{equation}
    \boldsymbol{\beta}_e(\mathbf r) =      
    \left[ \omega_{e_1}(\theta_{e_1})
    \frac{\hat{\boldsymbol{\rho}}_{e_1}}{\rho_{e_1}} \mathrm I_{A_{e_1}}(\mathbf r)
    -\omega_{e_2}(\theta_{e_2})
    \frac{\hat{\boldsymbol{\rho}}_{e_2}}{\rho_{e_2}} \mathrm I_{A_{e_2}}(\mathbf r)
    \right],
\end{equation}
where $\omega_{e_{1,2}}(\theta_{e_{1,2}})=\mu_e(\alpha) \, d\alpha/d\theta_{e_{1,2}}$ are the weight functions now in terms of the angles, $\rho$ and $\theta$ are the polar coordinates defined in Fig.~\ref{fig-2}-a, and $\text{I}_{A_{e_{1,2}}}$ are the indicator functions of the regions $A_{e_{1,2}}$. 
For an arbitrary contour used to compute the force on the vertex with Eq.~\eqref{eq-stress_force_vertex_stress_definition},  
the line element can be written as $\hat{\mathbf{n}}\,dS = \rho\, d\theta\, \hat{\boldsymbol{\rho}} - d\rho\, \hat{\boldsymbol{\theta}}$. With this, it is direct to verify that 
the above spread field gives the force on the vertex correctly and, therefore, the smoothed field $\boldsymbol{\sigma}^T(\mathbf{r})$ retrieves the correct microscopic forces in the vertex. Finally, the weights $\omega_{e_{1,2}}$ encode the fiber distributions emerging from the vertices. 
Note that they are not independent because for the fibers to match, $\theta_{e_1}$ and $\theta_{e_2}$ are related.

For the pressure component of the stress, inspired by the Murdoch--Hardy's approach~\cite{Barton2017,Admal2010}, we decompose the space in the cell areas $A_c$. Considering the isotropic nature of the pressure, we propose the pressure field 
\begin{equation}\label{eq-pressure_field}
    \boldsymbol{{\sigma}}^P(\mathbf{r})
    =
    -\sum_{c} \left ( P_c -\overline{P} \right ) {\text{I}}_{Ac}\left( \mathbf{r}\right)\mathbf{ \mathbb{I} },
\end{equation}
where the sum is over all cells and $\text{I}_{A_c}$ is the indicator function of the region $A_c$.
Here, $P_c - \overline P$ is the cell pressure, in reference to an average pressure $\overline{P}=\sum_m\lambda_mP_m$ to correct for the pressure degeneracy~\cite{Godolphim2025}, with $\sum_m\lambda_m=1$~\footnote[4]{Indeed, to obtain an expression free of degeneracy (that is, if $A_{0c}\rightarrow A_{0c}^{'}=A_{0c}+P_0/K_{Ac}$, $\Delta P_c$ is unchanged) it is enough to have $\Delta P_c = P_c - \sum_m\lambda_m^cP_m$ with $\sum_m \lambda_m^c=1$, but later, the correct force in Eq.~\eqref{eq.compute.fP} is only obtained if $\lambda_m^c=\lambda_m$.}.

As the proposed pressure field is uniform inside the cells, 
the force density  $\mathbf{q}^P(\mathbf{r})=\nabla \cdot \boldsymbol{{\sigma}}^P(\mathbf{r})$, is located only at the edges. 
Note that the force density is not equal to Eq.~\eqref{eq.forcedensity.qp}, which could not be written as a divergence due to the nonreciprocity of the pressure force. Hence, the proposed field smears out the microscopic force density over the edge.
Then, to apply Eq.~\eqref{eq-stress_force_vertex_stress_definition} for the pressure component of the force on the vertex and using that half of the force is allocated to each vertex [Eqs.~\eqref{force_edge_pressure_tension} or \eqref{eq.forcedensity.qp}], the contour $S_i$ must go through the edges centers, being otherwise immaterial its trajectory. A possible choice is to define the areas $A_i$ as the polygon that passes through all the cell centers and edge centers surrounding vertex $i$, as shown in Fig.~\ref{fig-2}-b. 
These regions tile the entire tissue without overlap and are consistent with the tilling produced by the sub-regions $A_{e_{1,2}}$, used to build the fiber bundles. Note that $A_i$ are similar, but not equal, to the regions used to calculate the tricellular-junction-stress in Refs.~\cite {Jensen2020,Jensen2023}.
Applying Eq.~\eqref{eq-stress_force_vertex_stress_definition} to $\boldsymbol{{\sigma}}^P(\mathbf{r})$ gives
\begin{align} \label{eq.compute.fP}
    \oint_{S_i} \boldsymbol{{\sigma}}^P(\mathbf{r})\cdot d\mathbf{s} 
    = 
    \sum_{c / i \in c}  P_c \frac{1}{2}\left ( \mathbf{r}_{i_c+1} - \mathbf{r}_{i_c-1} \right )\times \hat{\mathbf{z}},
\end{align}
which equals the expression for the pressure force in Eq.~\eqref{force_edge_pressure_tension}, regardless of the value of $\overline{P}$. The proposed expression~\eqref{eq-pressure_field} is degeneracy-free and, when integrated over any contour going through the edge centers, retrieves the correct microscopic pressure-forces in the vertex $i$. The degeneracy correction only introduces a constant field valued $\overline{P}$, which integrates to zero in any closed contour. The values of $\lambda_m$ are then irrelevant for the dynamics, but their values determine the spatial average of the field, hence fixing the tissue pressure gauge~\cite{Godolphim2025}: $\boldsymbol{\sigma}_\mathcal{T}^P \equiv \langle \boldsymbol{{\sigma}}^P(\mathbf{r}) \rangle_{A_\mathcal{T}} = \sum_{m} P_m \left ( \lambda_m - A_m/A_\mathcal{T} \right )\mathbb{I}$. For example,  the zero pressure gauge (ZPG) $\boldsymbol{\sigma}_\mathcal{T}^{P, \text{ ZPG}}=0$~\cite{Godolphim2025}, requires $\lambda_m^\text{ZPG} = A_m/A_\mathcal{T}$, resulting in $\boldsymbol{{\sigma}}^{P, \text{ ZPG}}(\mathbf{r}) = -\sum_{c} \left ( P_c - \langle P_c \rangle \right ) {\text{I}}_{A_c}\left( \mathbf{r}\right)\mathbf{ \mathbb{I} }$, where the spatial average of the pressure is $\langle P_c \rangle \equiv \sum_c A_c P_c/A_\mathcal{T}$. 

Since both $\boldsymbol{\sigma}^T(\mathbf{r})$ [Eq.~\eqref{eq-tension_field}] and $\boldsymbol{\sigma}^P(\mathbf{r})$ [Eq.~\eqref{eq-pressure_field}] can be integrated in common contours to give the vertex force $\mathbf{f}_i$, the tensorial stress field for the VM is then given by $\boldsymbol{\sigma}\left( \mathbf{r}\right) \equiv \boldsymbol{\sigma}^P\left( \mathbf{r}\right) + \boldsymbol{\sigma}^T\left( \mathbf{r} \right)$. This field is degeneracy-free, non-symmetric, defined everywhere in space, and, when integrated along any contour enclosing vertex $i$ that passes through the centers of the edges of $i$, it reproduces the correct microscopic forces at the vertex. 
For an open tissue, the pressure degeneracy is removed~\cite{Godolphim2025}, and it suffices to take $\overline{P}=0$ and the force on border vertices is obtained via the contour shown in Fig.~\ref{fig-2}-c subject to an external pressure $P^\text{ext}$,  while bulk forces are integrated as in the closed tissue case. 

The total force $\mathbf{F}_\mathcal{X}$ acting on a mesoscopic object $\mathcal{X}$ (see Supplemental Material) can be obtained from a suitable integration contour. For the case of a cell $c$, the total force $\mathbf{F}_c=\sum_{i/i\in c} \mathbf{f}_i$ is recovered for a contour taken arbitrarily close and outside the cell boundary, provided it passes through the centers of all external edges adjacent to $c$ (see Fig.~\ref{fig-2}-d and Supplemental Material). Lastly, torques can also be calculated for any mesoscopic object $\mathcal{X}$ (see Supplemental Material).

\textit{Coarse-grained expressions for the stress tensor.}
Computing the average in Eq.~\eqref{eq-stress_mesoscopi_stress_tensor_definition} over the whole tissue, a single cell, the edge area $A_e$ defined to build the fiber bundle (Fig.~\ref{fig-2}-a), and the vertex area $A_i$ used to compute the force due to the pressure (Fig.~\ref{fig-2}-b), the stress tensors read
(see Supplemental Material) 
\begin{align}
    \boldsymbol{\sigma}_\mathcal{T} &= - \left( \langle P_c  \rangle - \overline{P} \right) \mathbb{I} +
    \frac{1}{A_\mathcal{T}}\sum_e T_e l_e\hat{\mathbf{l}}_e\otimes\hat{\mathbf{l}}_e ,\\
    \boldsymbol{\sigma}_c &= - \left(P_c - \overline{P}\right)\mathbb{I}
    + \frac{1}{A_c} \sum_{e/c\in e} \Lambda_e^c T_e l_e\mathbf{\hat{l}}_e\otimes\hat{\mathbf{l}}_e,\\
    \boldsymbol{\sigma}_e  &= \frac{1}{A_e}\left[ -\left( P_e^\text{l} A_e^\text{l} + P_e^\text{r} A_e^\text{r} - A_e \overline{P}  \right)\mathbb{I}
    + T_e l_e\mathbf{\hat{l}}_e\otimes\hat{\mathbf{l}}_e \right],\\
    \boldsymbol{\sigma}_i   &= \frac{1}{A_i} \sum_{e/ i \in e} \left[ -\frac{1}{2}\left( P_e^\text{l} A_e^\text{l} + P_e^\text{r} A_e^\text{r}  -A_i \overline{P} \right)\mathbb{I} 
    + T_e l_e \hat{\mathbf{l}}_e \otimes \mathbf{m}_{e_i} \right].
\end{align}
Here,  
$\Lambda_e^c\equiv\int_{\mathbf{B}_e^\alpha\subset A_c} \mu_e(\alpha)d\alpha=\int_{\theta_1\in A_c} \omega_{e_1}(\theta_1) d\theta_1=\int_{\theta_2\in A_c} \omega_{e_2}(\theta_2) d\theta_2$ is the fraction of the filament bundle of edge $e$ that goes through cell $c$, 
$A_e^{\text{l,r}}$ are the cell areas at the left or right side of the edge $e$ (see Fig.~\ref{fig-2}-a), 
and $\mathbf{m}_{e_i} \equiv \int 
\boldsymbol{\rho}_{e_i}^{\max}(\theta) \omega_{e_i}(\theta)\,d\theta$ 
is the vertex \textit{tension geometric moment} in the region $A_e^i$, where $\boldsymbol{\rho}_{e_i}^{\max}$ is the vectorial length of each sub-filament (see Fig.~\ref{fig-2}-a).
Although $\mathbf{m}_{e_1} + \mathbf{m}_{e_2}=\mathbf{l}_e$, in general $\mathbf{m}_{e_i} \nparallel \mathbf{l}_e$. This implies that while $\boldsymbol{\sigma}_\mathcal{T}$, $\boldsymbol{\sigma}_c$, and $\boldsymbol{\sigma}_e$ are symmetric tensors, $\boldsymbol{\sigma}_i$ is not, and will be symmetric only for a regular tissue (similarly to what was reported in Refs.~\cite{Jensen2020,Jensen2023}), or provided that the filament bundle distributions  $\omega_{e_{1,2}}$ give  $\mathbf{m}_{e_{1,2}}=\mathbf{l}_e/2$. 

While $\boldsymbol{\sigma}_{\mathcal{T}}$ and $\boldsymbol{\sigma}_e$ are uniquely defined, with numerical values depending only on the geometry and model parameters, $\boldsymbol{\sigma}_c$ and $\boldsymbol{\sigma}_i$ explicitly depend on the filament distributions $\omega_{e_{1,2}}$. This dependence allows the recovery of $\boldsymbol{\sigma}_c^\text{Yang}$, obtained for $\Lambda_e^c=1/2$, meaning that edge forces are equally distributed between the two neighboring cells, and of $\boldsymbol{\sigma}_c^\text{virial}$, obtained for $\Lambda_e^c = K_{Lc}(L_c-L_{0c})/T_e$, meaning that edge-force contributions to the cellular stress arise entirely from the cell itself. In both cases, maintaining these stress representations throughout the tissue dynamics requires that the filament distributions $\omega_e^i$ adapt continuously over time to preserve the functional form of $\Lambda_e^c$. Conversely, for a fixed filament distribution, $\Lambda_e^c$ and $\mathbf{m}_{e_i}$ generally evolve with tissue geometry, leading to time-dependent changes in the functional form of $\boldsymbol{\sigma}_c$ and $\boldsymbol{\sigma}_i$.

\textit{Discussion.} The continuous stress field introduced here provides a link between microscopic VM forces and mesoscopic stress tensors. Within this framework, previously proposed VM stress tensors are not competing definitions of stress, but rather different coarse-grained realizations of the same underlying field. In this sense, the VM exhibits a stress non-uniqueness analogous to that found in molecular systems~\cite{IrvingKirkwood1950,soto2016kinetic}, where different stress constructions can arise from the same microscopic force field. The field also incorporates the pressure-gauge freedom of the VM, yielding a stress representation invariant under the parametric degeneracy~\cite{Godolphim2025}. 

As several force-inference approaches rely on VM-based stress representations, different choices of stress tensor can lead to different physical interpretations. The present framework provides a common reference from which these representations can be understood and compared. The edge-level stress tensor may be particularly useful in this context, as it is uniquely defined and retains higher spatial resolution than cell-level stresses.

We emphasize, however, that the construction presented here is not unique. The definition of the stress field has arbitrariness both in the pressure contribution and in how edge-tension forces are distributed through the tissue. Other constructions satisfying the same consistency conditions could, in principle, be formulated, leading to different stress fields and possibly different mesoscopic stress tensors. Within the present framework, tissue- and edge-level stress tensors are uniquely defined, whereas cell- and vertex-level stresses depend on how tension is distributed across each edge region through the functions $\omega_{e_{1,2}}$.

This raises the question of whether $\omega_{e_{1,2}}$ could be interpreted as an effective representation of cytoskeletal organization and stress fibers inside cells, with its time dependence linked to cytoskeletal remodeling and plasticity. If so, local stress measurements~\cite{Kong2019,Dow2023} or fluorescence microscopy of cytoskeletal networks could then be used to inform these distributions. The resulting stresses could be compared with independent experimental measurements, such as laser ablation or magnetic and oil droplets~\cite{Kong2019,Campas2014,Serwane2016,Lucio2017}, providing consistency checks for mechanical predictions and force-inference measurements.

\vspace{0.1cm}

\begin{acknowledgments}
\textit{Acknowledgments.} This research was supported by the Fondecyt Grants No.\ 1220536, 1260942, and NCN2024\_068 (SELFO) of ANID, Chile. P.C.G. expresses his gratitude for the grant No.\ 21231227 of ANID, Chile, the Brazilian agencies CAPES and FAPERGS for their financial support, and CNPq/PECI project number 200442/2026-0.
L.G.B. thanks CNPq for the grant 443517/2023-1.
We thank Silke Henkes for the interesting insights and encouragement, and Miguel Concha and the LEO-Scian group for the inspiring discussions.
\end{acknowledgments}

\section*{Supplemental Material}

\section{Vertex forces} 

The vertex variational force [Eq.~(1) in the main text] can be expressed as the sum of the forces $\mathbf{f}_i^c$ from the cells containing vertex $i$, or as the sum of the edge forces $\mathbf{f}_i^e$ acting on vertex $i$:
\begin{equation}\label{eq-stress_force_i_sum_c_sum_e}
\mathbf{f}_i
=
\sum_{c / i\in c} \mathbf{f}_i^c
=
\sum_{e / i\in e} \mathbf{f}_i^e
=
\sum_{e/i \in e}
\left(
\mathbf{f}_i^{e,P}
+
\mathbf{f}_i^{e,T}
\right),
\end{equation}
where the edge forces are given in the main text, and the cell forces read
\begin{align}\label{eq-stress_force_cell}
\mathbf{f}_i^c
&\equiv
-
\frac{\partial E_c}{\partial \mathbf{r}_i}
=
-\frac{1}{2}
P_c\,
\mathbf{r}_{i_c-1,i_c+1}
\times
\hat{\mathbf{z}}
\notag\\
&\quad
+
K_{Lc}
\left(
L_c-L_{0c}
\right)
\left(
\hat{\mathbf{r}}_{i_c,i_c+1}
+
\hat{\mathbf{r}}_{i_c,i_c-1}
\right),
\end{align}
where
\[
E_c=
\frac{1}{2}
\left[
K_{Ac}
\left(
A_c-A_{0c}
\right)^2
+
K_{Lc}
\left(
L_c-L_{0c}
\right)^2
\right]
\]
is the energy functional of a single cell, $\hat{\mathbf{z}}$ is the unit vector perpendicular to the tissue plane, and the vertex indices $i_c$ are cyclically ordered clockwise in cell $c$.

To obtain the right-hand side of Eq.~\eqref{eq-stress_force_cell},
the area of a planar polygonal cell can be computed by decomposing it into triangles as
$
A_c=
-
\sum_{j/j\in c}
\hat{\mathbf z}\cdot
(\mathbf r_j\times\mathbf r_{j+1})/2
$, where the sum runs over all the vertices in the cell. With this, only the terms containing vertex $i$ contribute to the derivative $-\partial /\partial \mathbf r_i
$, yielding
\begin{equation}
-\frac{\partial A_c}{\partial \mathbf r_i}
=
\frac12
\mathbf r_{i_c-1,i_c+1}
\times
\hat{\mathbf z}.
\end{equation}
Writing the perimeter as
$
L_c=
\sum_{j/j\in c}
|\mathbf r_{j+1}-\mathbf r_j|,
$
one obtains
\begin{equation}
-\frac{\partial L_c}{\partial \mathbf r_i}
=
\hat{\mathbf r}_{i_c,i_c+1}
+
\hat{\mathbf r}_{i_c,i_c-1}.
\end{equation}
Substituting these identities into
\begin{equation}
\mathbf{f}_i^c
=
-
K_{Ac}
\left(
A_c-A_{0c}
\right)
\frac{\partial A_c}{\partial \mathbf r_i}
-
K_{Lc}
\left(
L_c-L_{0c}
\right)
\frac{\partial L_c}{\partial \mathbf r_i},
\end{equation}
and using
$
P_c=-K_{Ac}(A_c-A_{0c}),
$
gives Eq.~\eqref{eq-stress_force_cell}.

The equivalence between the cell-force and edge-force representations in Eq.~\eqref{eq-stress_force_i_sum_c_sum_e} follows directly from summing the contributions of the cells sharing vertex $i$ and regrouping the resulting terms according to the edges incident on that vertex. Under this rearrangement, the area contributions combine into pressure-jump terms across each edge, while the perimeter contributions combine into the corresponding edge tensions, yielding the edge-force expression given in the main text. The same is also valid for boundary vertices, after setting the pressure and perimeter contributions of the missing neighboring cell to zero.

\section{Virial-like cellular stress tensor}

Virial stresses rigorously describe systems of interacting inertial particles in atomic and molecular systems~\cite{Admal2010}. Here, we introduce an analogous virial-like construction for the VM, without assuming \textit{a priori} its direct validity for this model. For the non-inertial VM dynamics, the kinetic contribution can be neglected. Retaining only the configurational term, the tissue-level virial-like stress reads $
\boldsymbol{\sigma}_{\mathcal{T}}^{\mathrm{virial}}
\equiv
\sum_i
\mathbf f_i
\otimes
\mathbf r_i/A_{\mathcal{T}}$,
where $A_{\mathcal T}$ is the tissue area (the use of absolute vertex positions follows from $E$ being translationally invariant). Cell-level virial-like stresses can then be obtained by decomposing the tissue stress using Eq.~\eqref{eq-stress_force_i_sum_c_sum_e}, yielding
\begin{align}
\boldsymbol{\sigma}_{\mathcal T}^{\mathrm{virial}}
&=
-\frac{1}{A_{\mathcal T}}
\sum_c
\sum_{i/i\in c}
\mathbf f_i^c
\otimes
\mathbf r_i
=
\frac{1}{A_{\mathcal T}}
\sum_c
A_c
\boldsymbol{\sigma}_c^{\mathrm{virial}},
\end{align}
with
\begin{equation}
\boldsymbol{\sigma}_c^{\mathrm{virial}}
\equiv
-\frac{1}{A_c}
\sum_{i/i\in c}
\mathbf f_i^c
\otimes
\mathbf r_i,
\end{equation}
where, again, the use of absolute position comes from the translational invariance of $E_c$.

Using Eq.~\eqref{eq-stress_force_cell}, $\boldsymbol{\sigma}_c^{\mathrm{virial}}$ can be split, for convenience, into a pressure component $\mathbf M$ and a tension component $\mathbf T$, so that
$\boldsymbol{\sigma}_c^\text{virial}
= \left(\mathbf{M}+\mathbf{T}\right)/A_c$.  For the pressure term,
\begin{equation}
\mathbf M
\equiv
-\frac{P_c}{2}
\sum_{i/i\in c}
\left(
\mathbf r_{i-1,i+1}\times\hat{\mathbf z}
\right)
\otimes
\mathbf r_i.
\end{equation}
Using cyclic summation over the polygon vertices and the polygon area expression
$
A_c
=
\sum_{i/i\in c}
\hat{\mathbf z}\cdot
(\mathbf r_i\times\mathbf r_{i+1})/2$,
one finds that the off-diagonal components cancel by cyclic summation, while the diagonal components satisfy
$M_{xx}=M_{yy}=-P_cA_c$. Hence
$
\mathbf M
=
-P_c\,A_c\,\mathbb I$.
For the tension term,
\begin{align}
\mathbf T
\equiv
K_{Lc}(L_c-L_{0c})
\sum_{i/i\in c}
\left(
\hat{\mathbf r}_{i,i+1}
+
\hat{\mathbf r}_{i,i-1}
\right)
\otimes
\mathbf r_i.
\end{align}
Relabeling indices in the second sum, combining both contributions, and using $\mathbf l_e=\mathbf r_{i+1}-\mathbf r_i$ and
$\hat{\mathbf r}_{i,i+1}=\mathbf l_e/l_e$, one gets $
\mathbf T
=
K_{Lc}(L_c-L_{0c})
\sum_{e/e\in c}
 l_e \hat{\mathbf{l}}_e \otimes \hat{\mathbf{l}}_e$.
Therefore, the virial-like cellular stress tensor reads
\begin{equation}
\boldsymbol{\sigma}_c^{\mathrm{virial}}
=
-P_c\,\mathbb I
+
\frac{K_{Lc}(L_c-L_{0c})}{A_c}
\sum_{e/e\in c}
l_e \hat{\mathbf{l}}_e \otimes \hat{\mathbf{l}}_e.
\end{equation}
The above expression have been proposed before in Refs.~\cite{Nestor-Bergmann2018,Perez-VerdugoThesis2021,Jensen2020,Jensen2023}. 

\section{Definition of forces and torques acting on mesoscopic objects}

The total force acting on a mesoscopic object $\mathcal{X}$ is the sum of all the forces acting in the vertices that belong to $\mathcal{X}$, $\mathbf{F}_\mathcal{X} \equiv \sum_{i\in \mathcal{X}} \mathbf{f}_i$. This is a necessary definition to correctly describe the equation of motion for the center of the object $\mathcal{X}$, which, for the overdamped dynamics, reads $\dot{\mathbf{R}}_\mathcal{X} =\mathbf{F}_\mathcal{X}$, where the center of viscosity $\mathbf{R}_\mathcal{X}$ equals the geometric center of $\mathcal{X}$. Also, the total torque acting on the center  of $\mathcal{X}$ reads $\boldsymbol{\tau}_\mathcal{X} \equiv \sum_{i \in \mathcal{X}} \mathbf{r}_i \times \mathbf{f}_i$.

Given the confluent nature of the VM, where vertices can be shared by more then one mesoscopic object, the total force $\mathbf{F}_\mathcal{X}$ can always be split into two contributions, $\mathbf{F}_\mathcal{X}=\mathbf{F}_\mathcal{X}^\text{self} + \mathbf{F}_\mathcal{X}^\text{neig}$, one coming from the energy function associated to the object $\mathcal{X}$ itself, and the other from the energy function associated to the neighbor objects $\mathcal{X}'$, which share vertices with $\mathcal{X}$. The same is true for the total torque $\boldsymbol{\tau}_\mathcal{X} = \boldsymbol{\tau}_\mathcal{X}^\text{self} + \boldsymbol{\tau}_\mathcal{X}^\text{neig}$. For a cell $c$, and considering its neighbor cells $c'$, the total force reads
\begin{equation}\label{eq-total_force_cell}
    \mathbf{F}_c = \sum_{i\in c}\mathbf{f}_i = \underbrace{\sum_{i\in c} \mathbf{f}_i^c}_{=0}  + \sum_{i\in c} \sum_{\substack{c'/i \in c'\\ c'\ne c}} \mathbf{f}_i^{c'},
\end{equation}
where $\sum_{i \in c}$ sums over all the vertices of cell $c$, and $\sum_{c'/i \in c',\,c'\ne c }$ sums over all neighbor cells of $c$ that are adjacent to vertex $i$. The vanishing term on the right-hand side of Eq.~\eqref{eq-total_force_cell} defines $\mathbf{F}_c^\text{self}$, which is always zero due to the translational invariance of $E_c$, while the non-vanishing term in the right-hand side defines $\mathbf{F}_c^\text{neig}$. Consequently, $\dot{\mathbf{R}}_c=\mathbf{F}_c^\text{neig}$.

\section{Calculation of forces and torques acting on a cell}

The precise value of $\mathbf{F}_c$ [Eq.~\eqref{eq-total_force_cell}] can be explicitly obtained using the VM equations, i.e., Eqs.~\eqref{eq-stress_force_i_sum_c_sum_e} and~\eqref{eq-stress_force_cell}, or, equivalently, as we now show by integrating the stress field $\boldsymbol{\sigma}(\mathbf{r})$ in a contour outside cell $c$ that passes trough the centers of all external adjacent edges, $S^\text{out}_c$, as the one shown in Fig.~2-d of the main text. 

\begin{figure}[htb]
    \centering
    \includegraphics[width=\linewidth,trim=25 477 25 40, clip]{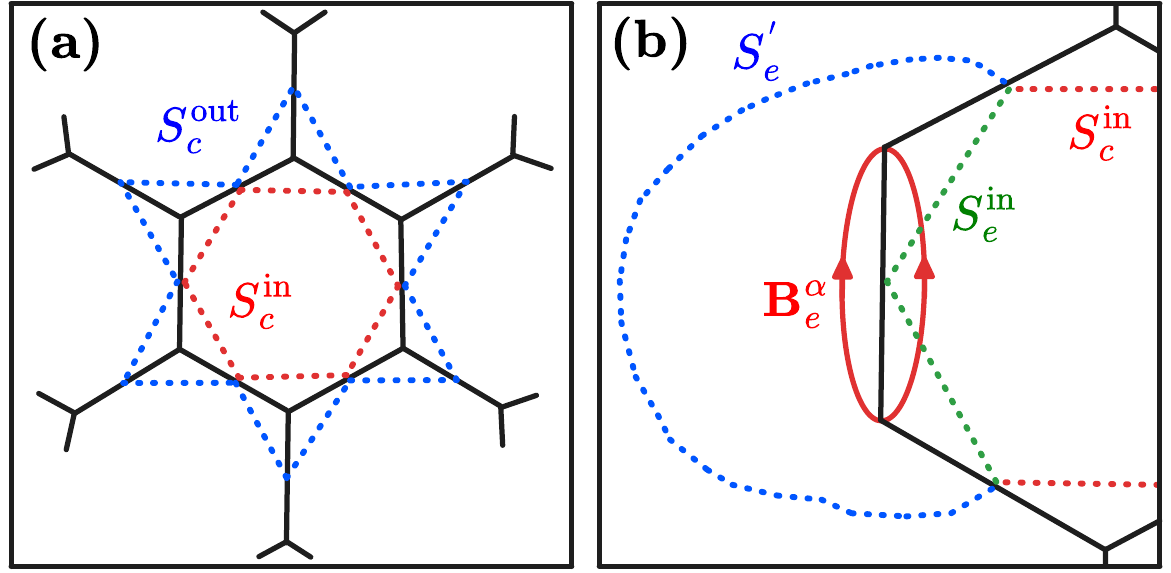}
    \caption{
    (a) Outer and inner integration contours associated with cell $c$. The outer contour $S_c^{\mathrm{out}}$ (blue) passes through the centers of the adjacent external edges and determines the total force acting on the cell. The inner contour $S_c^{\mathrm{in}}$ (red) encloses only the cell interior. The union of both contours defines closed loops around each vertex of the cell.
    (b) Contours used in the proof that the tension contribution of the inner contour vanishes. The contour $S_e^{\mathrm{in}}$ (green; a subpath of $S_c^\text{in}$) always intersects an even number of times the force-transmission fibers $\mathbf{B}_e^\alpha$ of edge $e$, whereas the auxiliary contour $S'_e$ (blue) completes the integration loop without intersecting any fiber. 
    }
    \label{fig-s1}
\end{figure}

To show that $\mathbf{F}_c^\text{out}\equiv \oint_{S^\text{out}_c} \boldsymbol{\sigma}(\mathbf{r})\cdot d \mathbf{s}$ equals $\mathbf{F}_c$  let us consider the outside integration contour $S^\text{out}_c$ shown in blue in Fig.~\ref{fig-s1}-a, which fulfills the condition explained above (and therefore it is equivalent to the contour in Fig.~2-d of the main text). For convenience, let us also consider the inner contour $S^\text{in}_c$ shown in the same figure. The sum of the two contours, $ S^\text{out}_c\cup S^\text{in}_c$, defines a series of closed triangular loops around each vertex in $c$. The integral over one closed triangular loop equals exactly the total force $\mathbf{f}_i$ acting on vertex $i$ (as explained in the main text). Hence, it  is direct that $ \oint_{S^\text{out}_c \cup S^\text{in}_c} \boldsymbol{\sigma}(\mathbf{r})\cdot d \mathbf{s}=\mathbf{F}_c$, which leads to  $\mathbf{F}_c^\text{out}=\mathbf{F}_c-\mathbf{F}_c^\text{in}$, where $\mathbf{F}_c^\text{in}\equiv \oint_{S^\text{in}_c} \boldsymbol{\sigma}(\mathbf{r})\cdot d \mathbf{s}$. It remains to show that $\mathbf{F}_c^\text{in}=0$. 

First, since the pressure field is constant inside the cell, the pressure contribution to the inner contour integration vanishes. For the tension part, we apply the reverse version of the Irving--Kirkwood procedure~\cite{IrvingKirkwood1950} used in the main text.
For that, let us consider an auxiliary path $S^{'}_e$ so that $S^\text{in}_e \cup S^{'}_e$ makes a closed loop around the edge $e$ (see Fig.~\ref{fig-s1}-b), and it does not intercept any of the force fibers of the edge $e$, implying that $\int_{S_e^\text{in}} \boldsymbol{\beta}_e(\mathbf{r}) \cdot d\mathbf{s}=\oint_{S_e^\text{in}\cup S_e^{'}} \boldsymbol{\beta}_e(\mathbf{r}) \cdot d\mathbf{s}$. With this, 
\begin{align}
    \mathbf{F}_c^\text{in} &= \sum_{e\in c}T_e \hat{\mathbf{l}}_e \oint_{S_e^\text{in}\cup S_e^{'}} \boldsymbol{\beta}_e(\mathbf{r})\cdot d\mathbf{s}\notag \\
    &= \sum_{e\in c}T_e \hat{\mathbf{l}}_e \oint_{S_e^\text{in}\cup S_e^{'}} \int d\alpha\, \mu_e(\alpha) \int_{\mathbf{B}_e^\alpha}\delta(\mathbf{r}-\mathbf{b})d\mathbf{b} \cdot d\mathbf{s} \notag \\
    &=  \sum_{e \in c} T_e \hat{\mathbf{l}}_e \int d \alpha \mu(\alpha) \int_{A_e^{'}} \int_{\mathbf{B}_e^\alpha}\nabla \delta(\mathbf{r}-\mathbf{b})d\mathbf{b}  d^2\mathbf{r} \notag \\
    &= - \sum_{e \in c} T_e \hat{\mathbf{l}}_e \int d \alpha \mu(\alpha) \int_{A_e^{'}} \left [ \delta(\mathbf{r} - \mathbf{r}_{e_2}) - \delta(\mathbf{r} - \mathbf{r}_{e_1}) \right ] d^2\mathbf{r} \notag \\
    &=0,
\end{align}
where, the divergence theorem is used in going from the second to the third line with $A_e^{'}$ is the area region enclosed by $S_e^\text{in}\cup S_e^{'}$, and from the third to the fourth line, we used that $\nabla_\mathbf{r} \delta( \mathbf{r} - \mathbf{b} )=-\nabla_\mathbf{b} \delta( \mathbf{r} - \mathbf{b} )$, which allows to perform the integral in $\mathbf{b}$. Finally, the last integral vanishes identically because both vertices $\mathbf{r}_{e_1}$ and $\mathbf{r}_{e_2}$ are always inside $A_e'$, which concludes the demonstration.

\section{Derivation of coarse-grained expressions for the stress tensor}

The VM stress field,
\begin{equation}
    \boldsymbol{\sigma}(\mathbf{r}) = -\sum_{c} \left ( P_c -\overline{P} \right ) \,{\text{I}}_{Ac}\left( \mathbf{r}\right)\,\mathbf{ \mathbb{I} } +  \sum_e  T_e\,\hat{\mathbf{l}}_e  \otimes  \boldsymbol{\beta}_e(\mathbf{r}),
\end{equation}
where
\begin{equation}\label{eq-SM_radial_spread_field}
    \boldsymbol{\beta}_e(\mathbf r) =      
    \left[ \omega_{e_1}(\theta_{e_1})
    \frac{\hat{\boldsymbol{\rho}}_{e_1}}{\rho_{e_1}} \mathrm I_{A_{e_1}}(\mathbf r)
    -\omega_{e_2}(\theta_{e_2})
    \frac{\hat{\boldsymbol{\rho}}_{e_2}}{\rho_{e_2}} \mathrm I_{A_{e_2}}(\mathbf r)
    \right],
\end{equation}
can be integrated to give coarse-grained averaged stress tensors, defined for the entire tissue or any mesoscopic region, using $ \boldsymbol{\sigma}_{\mathcal{X}} \equiv \left < \boldsymbol{\sigma}(\mathbf{r}) \right >_{A_\mathcal{X}} = \frac{1}{A_\mathcal{X}}\int_{A_\mathcal{X}} \boldsymbol{\sigma}(\mathbf{r})d^2\mathbf{r}$. For simplicity, here and in what follows, $A_\mathcal{X}$ represents both the geometrical region $\mathcal{X}$ and the numerical value of its area.
We consider the tissue, cell, edge, and vertex area regions $A_\mathcal{T}$, $A_c$, $A_e$, and $A_i$, respectively, as defined in the main text. 

Using the identities $\langle {\text{I}}_{A_\mathcal{Y}} \rangle_{A_\mathcal{X}}=A_\mathcal{Y}/A_\mathcal{X}$, for $A_\mathcal{Y}\subset A_\mathcal{X}$ and zero when $A_\mathcal{Y}\cap A_\mathcal{X}=\emptyset$,  $\langle \boldsymbol{\beta}_e \rangle_{A_\mathcal{X}}=l_e \hat{\mathbf{l}}_e/A_\mathcal{X}$, for $A_e\subset A_\mathcal{X}$ and zero when $A_e \cap A_\mathcal{X}=\emptyset$, and considering that the degeneracy correction $\overline{P}$ is homogeneous over the entire tissue, the tissue stress tensor directly reads 
\begin{align}
    \boldsymbol{\sigma}_\mathcal{T} \equiv \langle \boldsymbol{\sigma}(\mathbf{r}) \rangle_{A_\mathcal{T}} = - \left( \langle P_c  \rangle - \overline{P} \right) \mathbb{I} + \frac{1}{A_\mathcal{T}}\sum_e T_e l_e\hat{\mathbf{l}}_e\otimes\hat{\mathbf{l}}_e,
\end{align}
where $\langle P_c \rangle=\sum_c P_c A_c / A_\mathcal{T}$. 

For the cell stress tensor $\boldsymbol{\sigma}_c$, the cell $c$ area can be decomposed as $A_c = \bigcup_{e/e\in c}A_e\cap A_c$, where $A_e\cap A_c$ is the area region of cell $c$ inside $A_e$, so that $A_e = (A_e\cap A_c) \,\cup\, (A_e\cap A_{c'})$, where  $c$ and $c'$ are the cells that form the edge $e$. The spread field, integrated in the region $A_e\cap A_c$, then reads
\begin{align}
    \int_{A_e\cap A_c}\boldsymbol{\beta}_e(\mathbf{r})\,d\mathbf{r}^2
    =
    \int_{\Delta\theta_{e_1}^c}
    \hat{\boldsymbol{\rho}}_{e_1}(\theta_{e_1})\,
    \rho^{\max}_{e_1}(\theta_{e_1})\,
    \omega_{e_1}(\theta_{e_1})\,d\theta \notag \\
    -
    \int_{\Delta\theta_{e_2}^c}
    \hat{\boldsymbol{\rho}}_{e_2}(\theta_{e_2})\,
    \rho^{\max}_{e_2}(\theta_{e_2})\,
    \omega_{e_2}(\theta_{e_2})\,d\theta,
    \label{eq-branch_beta_start}
\end{align}
where $\Delta \theta_{e_{1,2}}^c$ are the angular domains of the region $A_e\cap{A_c}$ in the radial coordinate systems of vertices $e_1$ and $e_2$, and $\rho^\text{max}_{e_{1,2}}$ is the radial value of the intercept point in the interface line that cuts trough the region $A_e\cap A_c$, i.e., the straight line that connects the center of edge $e$ to the center of cell $c$. Given the bijectivity between $\theta_{e_1}$ and $\theta_{e_2}$, which relates to each  path $\alpha$ in the bundle  via $\omega_{e_{1,2}}(\theta_{e_{1,2}})=\mu_e(\alpha) \, d\alpha/d\theta_{e_{1,2}}$, that $\mathbf{l}_e =
\hat{\boldsymbol{\rho}}_{e_1}(\theta_\alpha)\,\rho^{\max}_{e_1}(\theta_{\alpha})
-
\hat{\boldsymbol{\rho}}_{e_2}(\theta_\alpha)\,\rho^{\max}_{e_2}(\theta_{\alpha})$, and that Eq.~\eqref{eq-branch_beta_start} equals $\mathbf{l}_e \Lambda_e^c$, where $\Lambda_e^c \equiv \int_{\alpha \in A_e\cap A_c} \mu_{e}(\alpha)\,d\alpha = \int_{\Delta_{e_{1,2}}^c} \omega_{e_{1,2}}(\theta_{e_{1,2}})\,d\theta$, then 
\begin{align}
        \boldsymbol{\sigma}_c = - \left(P_c - \overline{P}\right)\mathbb{I}
    + \frac{1}{A_c} \sum_{e/c\in e} T_e l_e\mathbf{\hat{l}}_e\otimes\hat{\mathbf{l}}_e \Lambda_e^c.
\end{align}

Given all of the above, it is direct that the edge stress tensor $\boldsymbol{\sigma}_e \equiv \langle \boldsymbol{\sigma}(\boldsymbol{r}) \rangle_{A_e}$ reads
\begin{equation}
    \boldsymbol{\sigma}_e = \frac{1}{A_e}\left[ -\left( P_e^l A_e^l + P_e^r A_e^r - A_e \overline{P}  \right)\mathbb{I}
    + T_e l_e\mathbf{\hat{l}}_e\otimes\hat{\mathbf{l}}_e \right],    
\end{equation}
where $A_e^{l,r}$ denote the area regions on the left and right sides of region $A_e$, and $P_e^{l,r}$ denote the cell pressures of the left and right cells at edge $e$. Finally, for the vertex stress tensor $\boldsymbol{\sigma}_i$, the vertex area is decomposed as $A_i=\bigcup_{e/ i \in e}A_e\cap A_i$, where, as above, $A_e=A_e^l \cap A_e^r$, but also, as defined in the main text, $A_e = A_{e_1}\cup A_{e_2}$, $A_{e_1}\cap A_{e_2}=\emptyset$, where $A_{e_{1,2}}$ are the vertex area regions in $A_e$, and they are defined so that for the area values $A_{e_1}=A_{e_2}=A_e/2$. Given all that, it is direct that the vertex stress tensor reads
\begin{align}
        \boldsymbol{\sigma}_i  = 
        \frac{1}{A_i} \sum_{e/ i \in e}\!\! \left[ -\frac{1}{2}\left( P_e^l A_e^l + P_e^r A_e^r  -A_i \overline{P} \right)\mathbb{I} 
    + T_e l_e \hat{\mathbf{l}}_e \otimes \mathbf{m}_{e_i} \right],
\end{align}
where the vertex $i$ correspond to vertex $e_1$ of the edge $e$ with edge vector $\mathbf{l}_e=\mathbf{r}_{e_2}-\mathbf{r}_{e_1}$, and $\mathbf{m}_e^i \equiv \int 
\boldsymbol{\rho}_{e_i}^{\max}(\theta_{e_i})
\omega_{e_i}(\theta_{e_i})\,d\theta$ integrates over the entire natural domain of $\theta_{e_i}$.

\end{document}